\documentstyle[12pt]{article}
\def\eqi{\begin{equation}}
\def\eqj{\end{equation}}
\def\etal{{\sl et al.\/}}

\def\pl{{\sl Phys.~Lett.\ }}
\def\pr{{\sl Phys.~Rev.\ }}

\def\Zf{{\sl Z.~Physik\ }}
\def\v#1{{\bf #1\ }}

\def\MeV{\hbox{{\rm MeV}}}

\def\Bbar#1{\mkern 0mu\overline{\mkern 0mu {#1}
           \mkern -1mu}\mkern 1mu}
\def\bbar#1{\mkern 0mu\overline{\mkern 0mu {#1}
           \mkern -2mu}\mkern 2mu}

\def\decale#1{\par\noindent\hskip
           3em\llap{{\it #1}\enspace}\ignorespaces}

\def\gal{\ {\raise .2em\hbox{$>$}}\hskip -.8em{\raise
             -.2em\hbox{$<$}}\ }
\def\gappeq{\mathrel{\rlap{\raise
           .5ex\hbox{$>$}}{\lower .5ex\hbox{$\sim$}}}}
\def\ie{i.e.\ }

\def\ket#1{\left|{#1}\right\rangle}
\def\lag{\ {\raise .20em\hbox{$<$}}\hskip -.8em{\raise
               -.20em\hbox{$>$}}\ }
\def\lappeq{\mathrel{\rlap{\raise .5ex\hbox{$<$}}
                          {\lower .5ex\hbox{$\sim$}}}}

\def\NN{{\rm N}\mkern -2mu{\rm N}}
\def\NNb{\PN\!\PBN}

\def\SLJ#1#2#3{^{#1}{\rm #2}_{#3}}

\def\phm{\ifmmode\phantom{-}\else\phantom{--}\fi}

\def\ppb{\Pp\mkern -1mu\Pbp}

\def\jipsi{{\rm J}\!/\Psi}
\def\P#1{{{\rm #1}} }
\def\Pe{{\rm e}}

\def\Pp{{\rm p}}
\def\PN{{\rm N}}
\def\PK{{\rm K}}

\def\PD{{\rm D}}
\def\PQ{{\rm Q}}
\def\Pq{{\rm q}}
\def\Pu{{\rm u}}
\def\Pd{{\rm d}}
\def\Ps{{\rm s}}
\def\Pc{{\rm c}}

\def\Pb#1{\bbar{{\rm #1}}}
\def\PB#1{\Bbar{{\rm #1}}}

\def\Pbp{\bbar{{\rm p}}}
\def\Pbq{\bbar{{\rm q}}}
\def\Pbu{\bbar{{\rm u}}}
\def\Pbd{\bbar{{\rm d}}}
\def\Pbs{\bbar{{\rm s}}}
\def\Pbc{\bbar{{\rm c}}}
\def\PBN{\Bbar{{\rm N}}}

\def\PBK{\Bbar{{\rm K}}}
\begin{document}
\begin{titlepage}
\begin{flushright}
\par\vglue -2cm
ISN 93-17
\end{flushright}
\vfill
\begin{center}
{\Large\bf SPIN AND FLAVOUR:}
\vskip .3cm
{\Large\bf  CONCLUDING REMARKS
\footnote{Invited talk at the Workshop on Spin and Flavour in Hadronic
and Electromagnetic Interactions,
Turin, September 1992, to appear in the Proceedings}}\\
\vglue 1cm
{\bf Jean-Marc Richard}\\
\vskip .1cm
{\small Institut des Sciences Nucl\'eaires}\\
{\small Universit\'e Joseph Fourier--CNRS--IN2P3}\\
{\small 53, avenue des Martyrs, F--38026 Grenoble, France}\\
\vglue 1.5cm
{\bf Abstract}\\
\vglue.1cm
\end{center}
We review some of the salient results presented at this
Workshop, together with some comments on the underlying physics, and
the proposed facilities for future experiments.
\par  \vfill \begin{flushleft} ISN 93--17\\  \today. \end{flushleft}
\end{titlepage}
\vglue .2cm
\begin{center}
{\Large\bf SPIN AND FLAVOUR:}
\vskip .25cm
{\Large\bf  CONCLUDING REMARKS}
\vskip 1.cm
{\bf Jean-Marc Richard}\\
\vskip .5cm
{ Institut des Sciences Nucl\'eaires}\\
{ Universit\'e Joseph Fourier--CNRS--IN2P3} \\
{ 53, avenue des Martyrs, F--38026 Grenoble, France}\\
\vglue 1.cm

{\bf Abstract.}
\end{center}
\parindent=0cm
 We review some of the salient results presented at this
Workshop, together with some comments on the underlying physics, and
the proposed facilities for future experiments.
\par
\vglue .2cm
\parindent=.4cm
\section{Introduction}\label{Intro} The Workshop
on {\sl Spin and Flavour in Hadronic and Electromagnetic
Interactions\/} was organized at Torino by R.~Bertini, F. Balestra and
R.~Garfagnini,
with the efficient help of their colleagues and secretaries.
It gave the opportunity of making interesting comparisons between
high-energy and low-energy problems, hadronic
and photon beams, old and new physics.\par
We have learned, or have been reminded, that there are intriguing
spin effects in high-energy hadronic collisions, and this seems to

challenge Quantum Chromodynamics (QCD) or at least its perturbative version
(pQCD). Striking spin effects also show up at very low energy, for
instance in the reaction $\ppb\rightarrow\Lambda\PB{\Lambda}$, and here,
one is tempted by a more conventional type of picture, involving
hadron-exchange between hadrons.\par
Many important  results in the field of this
Workshop come from LEAR and the $\ppb$ experiment E760, while electron or
photon beams are involved in projects like DA$\Phi$NE or GRAL which have
been presented.\par
In the course of the discussions, especially during the theory talks, I
was amazed by the mixing of modern concepts, such as
pQCD, Skyrmions, etc., and more conventional approaches
based on meson exchanges, potential models or Regge trajectories.
One of the first speakers quoted a great physicist, who did much
pionnering work on strong interactions, but might have
understimated the role of spin forces at high energy. Let me, in turn,
recall a remark made once by the same physicist on the $\Delta$
resonance, which is described either as a $\Pq^3$ cluster or as a
$\pi\PN$ state. Both pictures are probably valid. The
former is usually more appropriate in the general framework of hadron
spectroscopy, while the later is more efficient for describing
the $\Delta$ inside a nucleus.\par The equivalence between quark and
hadron pictures is the basis of the ``QCD sum rules'', an approach to
strong interactions which has proved rather successful. It includes an
interesting mixture of old-fashioned
 dispersion relations and modern field theory. The Skyrmion model is
also based on the possible equivalence between QCD, initially written
in terms of interacting fermions, and a reformulation in terms of
bosons. The equivalence remains however to be demonstrated with the
actual number of colours and dimensions.\par
There is also a duality in the presentation of the experimental
projects. Shooting photons, antiprotons or kaons on a nucleon
or a nucleus can clearly not be considered as a new experiment. The
promotors stress together the need for improved statistics, to answer
old questions, and the possibility of looking at new observables, to
test the most recent speculations.\par
I shall shortly comment on what we have heard on strangeness, charm,
and spin physics, and then on the approved or proposed experimental
facilities. This will hardly cover all the subjects which have been
discussed. In particular, I will not comment much  on the interesting
discussions on neutrino scattering \cite{Neutrino},
hadron dynamics in the nuclear medium \cite{Hadrons_in_nucleus}, and
light meson physics~\cite{Peaslee}.
\section{Strangeness}
We had a variety of talks dealing with particles containing
strange quarks:  mesons with hidden
 strangeness, hyperons and hypernuclei with increasing
strangeness. States with both charm and strangeness will
be discussed in the next section.
\subsection{Hidden strangeness}
The quark picture of mesons is usually considered as being under
control, especially in the sector of vectors ($J^P=1^-$), where one
registers only a small departure from the ideal mixing of
$$(\Ps\Pbs)\qquad\hbox{and}\qquad\left({\Pu\Pbu+\Pd\Pbd\over\sqrt2}
\right)_{I=0}.$$ However, recent LEAR data~\cite{Amsler,Bressani} show
that ratios of the type
$${\ppb\rightarrow\Phi+X\over\ppb\rightarrow\omega+X}$$ are together
larger than expected and state-dependent. This raises questions about
mechanisms for violation of the OZI rule  and in particular, the
influence of multiquark resonances on the dynamics of annihilation.\par
I note that the $\Phi\Phi$ channel is presently studied at LEAR,
in a narrow $\sqrt s$ range, which could be enlarged if
SuperLEAR is built. Studies on electro- or
photo-production of the $\Phi$ were also presented by Laget
\cite{Laget}, who reviewed several aspects of strangeness
production.\par
Radiative decays of the $\Phi$, such as $\Phi\rightarrow\gamma+{\rm
S}$, where S is a scalar ($J^P=0^+$) meson, can reveal some aspects of the
$\Phi$
wave function, but this is so far more oriented toward studying the
scalar mesons themselves, whose $\Pq\Pbq$ vs.\ $(\Pq)^2(\Pbq)^2$
vs.\ $\PK\PBK$ nature is highly controversial~\cite{Kumano}.\par
The situation is presumably more delicate in the pseudoscalar sector
than in the vector one. Here, besides the $(\Ps\Pbs)$ and
$(\Pu\Pbu+\Pd\Pbd)$ configurations, one expects a rather large gluonic
component, especially for the $\eta'$ meson. This is why looking at the
relative importance of $\eta\eta$, $\eta\eta'$, and $\pi^0\pi^0$
decays is acknowledged as crucial for separating glueballs from
multiquarks.\par
There are many new data on the decays of neutrals into $\eta$, $\eta'$
and $\pi^0$, thanks to the Cristal Barrel experiment at
LEAR~\cite{Amsler}. We have also been reminded that $\eta$ and $\eta'$
production is studied at Saturne~\cite{Dellacasa}. To the extent that
$\eta$ are produced through specific $\PN^*$ resonances, on can
study how these $\PN^*$ behave in the nuclei.\par Not surprisingly, one
thinks of building $\Phi$ or $\eta$ factories, to study CP violation
and rare, very rare or forbidden decays~\cite{Piccolo}. These are very
delicate experiments, which compete with those based on high-energy
kaon beams. \subsection{$S=-1$ baryons and hypernuclei}
We have heard that progress have been made recently or can be expected
on the magnetic moments of hyperons \cite{Lach}. This helps measuring
small departures of the naive quark model, which works astonishingly
well.\par
The data clearly show that the masses and magnetic moments of
ground-state baryons exhibit a very smooth behaviour in flavour space.
For instance, we have for the mass differences
$\Lambda-\PN <\Delta-\PN$, meaning that flavour excitation costs less
than a single spin excitation. This is rather naturally explained in
the very naive constituent models with flavour-independent confinement
and moderate quark-mass difference $m_\Ps-m_\Pq$.\par
On the other hand, one needs a more laborious tuning in some more
ambitious models, where at zeroth order, spin is averaged and flavour
excitation is pushed very high.\par
The excitations of hyperons are either not too well known, or poorly
understood by simple extrapolation of the non-strange sector. We are
all aware of the long-standing problem of the
$\Lambda(1520)-\Lambda(1415)$ splitting. There are
difficulties with other multiplets.
Moreover, several states in the $\Lambda$
 or $\Sigma$ spectrum await confirmation or discovery,
and this is a severe handicap for studying how the strange quark feels
confinement forces.\par
Hypernuclei with $S=-1$ have been known for years, thanks to $(\PK^-,\pi^-)$
and other experiments. This was reviewed by Gal~\cite{Gal}. An
interesting progress was made at CERN, with heavy hypernuclei seen as
delayed fission products following $\Pbp$ annihilation. Refined
measurements at KEK, in the $^{12}$C sector, includes polarization and
a comparison of mesonic vs.\ non-mesonic decay modes~\cite{Noumi}.\par
\subsection{$S=-2$ dibaryons and hypernuclei}
As recalled by Gal (see the references in his paper~\cite{Gal}), the H
(uuddss) was first predicted by Jaffe in 1977, with a simple wave
function
$$\ket{H}=\sqrt{{1\over5}}\left(\sqrt{1\over8}\ket{\Lambda\Lambda}
+\sqrt{4\over8}\ket{\Xi\PN}
+\sqrt{3\over8}\ket{\Sigma\Sigma}\right)+\cdots$$
where the dots can be read
either as an hidden-colour component or as  the result of permuting the
quarks. This wave function optimizes the chromomagnetic interaction and
leads to a sizeable binding of $\sim 150\,$MeV. The stability suffers
however from many types of effects: chromoelectric forces, kinetic
energy, breaking of the SU(3) symmetry, etc. As a result, the predictions
 considerably vary from one author to another, and even the
existence of the H is controversial.\par The H has been sought for in
several experiments, without success so far. The problem is that the
behaviour of the hypothetical H depends dramatically on its mass. One
does not design the same detector for a $\Lambda\Lambda$ resonance, a
loosely bound state which decays weakly, and a very stable state to be
seen from its  missing mass and momentum.\par
An indirect indication against the H is provided by
the $S=-2$, $B>2$ hypernuclei. Comparing the binding
energy with those of $S=-1$ hypernuclei leads
to an estimate
$$\delta B_{\Lambda\Lambda}\equiv B_{\Lambda\Lambda}-2B_\Lambda
\simeq4-5\,\MeV,$$
which can be interpreted in terms of average
baryon--baryon potentials (attractive) as
$$\mid V_{\Lambda \PN}\mid<\mid V_{\Lambda \Lambda}\mid<
\mid V_{\PN\PN}\mid,$$
The later inegality makes it unlikely for a
$\Lambda\Lambda$ bound state to exist, unless there is some very short
range attraction separated by a repulsive
barrier from the moderate attraction seen in $S=-2$ hypernuclei.
New experiments are planned, with stopped $\Xi$, $(\PK^-,\PK^+)$
reactions, or with the strange quarks produced in $\Pbp$ annihilation.
\subsection{Larger $(-S)$ states}
We have seen that the speculations  on strange
matter first written in terms of quarks  by Fahri, Jaffe, etc.\ have
been reanalyzed using a more conventional baryon basis~\cite{Gal}.
The results are qualitatively similar, and very
dramatic. When one tries to increase the average strangeness
$(-S)/A$, one better spreads strangeness into different types
of hyperons. Then $\Sigma$'s, $\Xi$'s and heavier hyperons supplement
the $\Lambda$'s as constituents of the nucleus, and the resulting
hypernucleus becomes more and more stable. This new spectroscopy

 is unfortunately out of reach of our experimental devices.
\section{Charm}
\subsection{Hidden charm}
Our knowledge of Charmonium is now much more accurate, thanks to the
measurements of the E760 experiment at Fermilab, which confirm and
improve the results obtained earlier by the R704 collaboration at CERN.
The $\ppb$ reaction gives access to all quantum numbers and allow
for very precise mass determinations.\par
It was underlined by Palestini~\cite{Palestini} that the mass of the
$\SLJ1P1$ state almost coincides with the centre-of-gravity of the
triplet states, namely
$$\SLJ1P1=\left[{1\over9}(\SLJ3P0)+{3\over9}(\SLJ3P1)+
{5\over9}(\SLJ3P2)\right]+1\,\MeV.$$
This confirms the most accepted ideas about spin forces: the spin-spin
term is very short ranged, and thus does not affect the $\ell>0$
partial waves, whose wave function vanishes at the origin.\par
 One would
be tempted to jump on the small ``$1\,\MeV$'' departure and interpret it
in terms of $\alpha_\Ps^2$ corrections in QCD (running coupling constant
and vertex corrections), smearing of the Breit--Fermi contact term,
relativistic effects, deviations from a purely scalar confinement, etc.
However, at this level of precision, one should first account for more
trivial effects. Let me quote some of these: \decale{1)} The
centre-of-gravity formula (see above) eliminates spin-orbit and tensor
forces only at first order. In the Hamiltonian, the spin-orbit and
tensor couplings enter linearly and vary from one $\SLJ3PJ$ state
to another. The lowest eigenvalue is thus  a concave function of
these couplings \cite{Thirring}. This means that the fictitious triplet
state free of spin-orbit and tensor forces should lie slightly {\sl
above} the naive centre of gravity. My numerical estimates give
something like $3-5\,\MeV$, indicating that the spin--spin force is
{\sl attractive}  in the $\ell=1$ sector. \decale{2)} From  general
symmetry considerations (and the literature on $\NN$, $\NNb$, etc.) we
know that a spin 1/2 -- spin 1/2 system as $\Pc\Pbc$ involves {\sl five}
independent spin operators. A type of ``quadratic spin-orbit'' operator
should supplement the usual central, spin-spin, spin-orbit and tensor
terms.  \decale{3)} The tensor force induces a small mixing between
$\SLJ3P2$ and $\SLJ3F2$, which slightly lowers the former.\par
In short, the phenomenological analysis becomes rather delicate at
the MeV level, and probably requires a relativistic framework.\par
New data are expected on the $\eta'_\Pc$ whose present status is
rather weak. It was noted by many authors (see, e.g., Ref.~\cite{MR}),
that the ratio
$${\Psi'-\eta'_\Pc\over\jipsi-\eta_\Pc}={93\,\MeV\over112\,\MeV},$$
as given by $\Pe^+\Pe^-$ data,
is a little too large to be easily reproduced in potential models.
With the recent increase of the hyperfine splitting
($\jipsi-\eta_\Pc$) of the ground state in $\Pbp$ data, this becomes
even more problematic.\par It was noted in \cite{MR} that some
sophisticated models with explicit account for the coupling to
continuum states,
$$\ket{\Psi}=\ket{\Pc\Pbc}+\epsilon\ket{\Pc\Pbq,\Pbc\Pq}+\cdots$$ do
not improve the fit of fine and hyperfine splittings. This is an
amazing paradox that the simplest models work at best.\par There are
also open questions concerning the decay of Charmonium. Most of them
are understood in terms of  a simple mechanism folded with the
probability $|\Phi(0)|^2$ of finding the quark and the antiquark
together. However, the ratio
$${\Psi'\rightarrow\pi\rho\over\jipsi\rightarrow\pi\rho}$$ seems
abnormally small, as if a mysterious long-range component would make it
sensitive to the node structure, or as if it was influenced by a resonance
sitting near the $\jipsi$.\par Anselmino \cite{Ansel}  also pointed out
that the branching ratio for $$\eta_\Pc\rightarrow\ppb$$ is much larger
than the most optimistic theoretical expectations. This is fortunate
for experimentalists, and this might help the projects of doing refined
$\Pc\Pbc$ or even $\P{b}\Pb{b}$ spectroscopy with antiprotons
(SuperLEAR project~\cite{Amsler}). This is however a challenge for
theorists. I note that in a similar energy range, one starts measuring
branching ratios for flavoured mesons (with b or c) into
baryon--antibaryon pairs: they are due to weak forces, but the
hadronization might be comparable.\par As noted by Amsler
\cite{Amsler}, SuperLEAR is not just a machine to do quarkonium
spectroscopy. This is a good place to look at hybrid states, of content
$\Pc\Pbc g$. Those states can be seen as excitations of the gluon field
surrounding the heavy quarks. Hybrids might exist in several sectors of
hadron spectroscopy, but they are better seen in Charmonium: the
spectrum of ordinary $\Pc\Pbc$ excitations is well measured and well
understood, so any exotic state that does fit as a radial or orbital
excitation is easily singled out. \subsection{Open charm}
The spectroscopy of D mesons is regularly improved, and some orbital
excitations are now identified~\cite{PDG}. During recent months, more
activity was devoted to charmed baryons. The subject was reviewed by
Paul \cite{Paul}. Again, simple potential models, extrapolated from
ordinary and strange baryons, account quite reasonably for the masses
of charmed baryons with or without strangeness. I notice in a recent
paper by Riska~\cite{Riska}, that his estimates of Qqq masses in the
Skyrmion model are significantly lower than these predicted by potential
models.\par
Besides spectroscopy,  charmed baryons
 are interesting for studying the weak decays and
subsequent hadronization. From the comparison between $\PD^+$ and
$\PD^0$ lifetimes and branching ratios, we understand that the
charmed quark, while decaying, does not ignore its environment. The
same is true for baryons, and one expects significant differences
between the decay patterns of cud, csu, csd, and css.\par
There are many developments one might dream of in the field of charmed
baryons: magnetic moments, detailed decay properties, orbital
excitations, etc. One can anticipate a stimulating competition between
various experiments with electron or hadron beams.
\par
The sector of baryons with two heavy quarks, previously considered as
a pure speculation~\cite{Fleck}, is now taken more seriously: there is
some hope to produce them at Fermilab~\cite{Quigg}, and there are
presently new studies on their spectroscopy ~\cite{Narison} and decay
properties~\cite{Quigg,Narison}. If QQq baryons are accessible, one
should also look at $\PQ\PQ\Pbq\Pbq$ mesons, whose existence has been
predicted~\cite{Zouzou}, as a consequence of flavour independence.
Baryons with charm $C=1$ and strangeness $S=-1$ have been
seen~\cite{PDG}, and are under active study at CERN and
Fermilab~\cite{Paul}. There are speculations about baryons with $C=-1$
and $S=-1$ or $-2$, \ie  of quark structure [$\Pbc$suud] or
[$\Pbc$ssud]~\cite{Penta,Riska}. There are active searches, again at
both CERN and Fermilab~\cite{Paul,Buenerd}.
\subsection{Charm in nuclei} The subject was discussed by Seth
\cite{Seth} and also touched by other speakers.\par
One might first think of charmed hypernuclei. Since $\Lambda_\Pc$ is
heavy, it would experience the inner part of the nucleus. But these
states are not easily produced. The kinematics is much less favourable
than for ordinary hypernuclei.\par
More fashionable is the $\Pc\Pbc$--nucleon or $\Pc\Pbc$--nucleus
interaction. In principle, this is a clean example of interhadronic
force without quark interchange, so a nice laboratory for studying the
Van-der-Waals regime of {QCD}. It may be that the interaction is attractive
enough to
produce
$\Pc\Pbc$--nucleus bound states.\par
The so-called $\jipsi$ suppression is considered as a signature for the
quark--gluon plasma. The interpretation of the data coming from
relativistic heavy ions requires the knowledge of ``ordinary''
rescattering effects, \ie the $\jipsi$ and $\Psi'$ cross section on
nucleons and nuclei.\par
One usually accepts the idea that these cross-sections are small,
because Charmonia are small objects, and that
$\sigma(\Psi')>\sigma(\jipsi)$. In fact  $\jipsi$ and $\Psi'$ have two
radii: one is the mean $\Pc\Pbc$ separation, which is indeed small
for $\Psi'$, and even smaller for $\jipsi$; the second is the
radius of the gluon field surrounding the quarks, typically $1\,$fm in
bag models, for both $\Psi'$ and $\jipsi$. So, if one believes that
the gluons contribute  to the cross-section, one expects
$\sigma(\Psi')$  and $\sigma(\jipsi)$ to be nearly equal and not too
small.\par
Studying how charmed quarks are produced in nuclei is one of the
goals of future high-intensity machines.
 \section{Spin}
Spin observables were mentioned in almost every talk,
 including these
on future facilities. I shall come back only on three topics: $\NNb$
scattering, $\Lambda$ production and high-energy reactions.
\subsection{$\NNb$ scattering}
The experimental results and their interpretation were reviewed by
Bradamante \cite{Bradam}. The $\NNb$ interaction is very strong and
includes many contributions: meson exchanges, $s$-channel resonances,
annihilation. To learn something, one has to apply filters, which
enhance one component after the other. This is precisely the job
of spin parameters, and already constraints have been set on the
phenomenological models from the data on analysing power and
depolarization.\par
If I had to comment on the theoritical activity, I would say
that one does not gain much by searching the minimal $\chi^2$ in
models with many parameters. Sometimes, after many hours of
expensive computing, one is not able to say which ingredients
 of the model are crucial,
and which new observables are worth measuring.\par
 One gets better insight by crude fits
with simple models whose physical content is better understood.
For instance, one expects that the combined  contributions of
pseudoscalar and vector meson exchanges lead to dramatic
tensor forces. This induces  large values of specific rank-2
observables, with longitudinal polarization, whose measurement
has never been done. It is a pity that the LEAR programme of
$\NNb$ scattering has been stopped. I hope it will be resumed
very soon.\par As reviewed by Amsler \cite{Amsler} and Bressani
\cite{Bressani}, we have now many data on the branching ratios for
annihilation at rest. One cannot understand annihilation without a
good control of the initial state interaction, which is strongly
spin and isospin dependent. This is why the annihilation and the
scattering experiments with antiprotons are complementary.\par
This complementarity is nicely illustrated by the data

of the PS172 collaboration on the $\ppb\rightarrow\pi\pi$ and
$\ppb\rightarrow\PK\PBK$ reactions. They found very large
asymmetry parameters, nearly saturating the unitarity limit
 in a wide angle and energy range. This implies a
maximal interference between the two independent helicity
amplitudes $F_{++}$ and $F_{+-}$, which is due to the strong
tensor force in the initial state, and the non-local character
of the annihilation operator~\cite{Elchikh}.

 \subsection{$\Lambda$ production}
The hyperons produced in  high-energy experiments
are polarized, even at very large momentum transfer, while
antihyperons produced on nuclear targets are not polarized
\cite{Lach,Ansel,Barni}.
\par
At first sight, this seems a leading particle effect associated
with an intrinsic polarization of the strange quark.
If, indeed, one identifies the spin of the $\Lambda$ with that
of the strange quark, the spin of the $\Sigma$ with its opposite, and
the spin of the $\Xi$ with half the spin of the ss pair, then a $20\%$
polarization of the strange quark explains the data on hyperons.\par
The problem is that one does not believe
anymore that the spin of a baryon
is so simply related to the spins of its valence quarks:
recent experiments on lepton scattering
have taught us to be cautious, though the situation might
depend on the region of the structure functions one is looking
at.\par The study of $\Lambda$ polarization is now performed at
lower energy. An experiment has been approved at Saturne
\cite{Bertini}, based on the reaction
$\Pp\Pp\rightarrow\PN\PK\Lambda$. The comparison with

$\Pbp\Pp\rightarrow\Pbp\PK\Lambda+{\rm cc.}$ (feasible at CERN)
would be instructive.\par The PS185 collaboration at CERN has
beautiful data on the hypercharge-exchange reaction
$\ppb\rightarrow\Lambda\PB{\Lambda}$: cross-section, polarization
and spin correlation in the final state. An intriguing result is
that the reaction takes place always  in the triplet state,
instead of $75\%$ of the time only if spin-dependent
forces would be absent. One has difficulties understanding  why the
transition is so much suppressed in the spin singlet case.
\subsection{QCD and spin effects} This is a well-known and
well-ignored problem. Some physicists persistingly stress the
failure of QCD, or at least of the naive approach to QCD,
for the spin effects which are observed at large energy and
momentum transfer. The rest of the community, unfortunately, does
not care too much about the spin measurements, and behaves as if
everything was fine with QCD. At this point, this becomes more a
problem of sociology than of physics.\par
Some the hotest questions have been reviewed by
Anselmino~\cite{Ansel} and also discussed by
Maggiora~\cite{Maggiora}. Among the possible remedies, Anselmino suggested
the use of diquark clusters for constructing the baryon wave
functions. Diquarks will be discussed at length at another
Workshop, which will also take place at Villa Gualino. \par
{}From what we heard on the surprizing results of spin measurements,
we should seriously study the possibility
of polarizing the beams, when designing the future
accelerators. Otherwise, one risks missing important pieces of
physics.\par
The same is true for electro-weak physics. To test the standard
model
in detail, and look at possible departures, for instance a
restoration of left-right symmetry, one has to use spin
measurements. \section{Future facilities}
Everyone in the audience has in mind his favourite project, for
continuing the physics we have discussed along this
Workshop: kaon factory (EHF, KAON), $\tau$--charm factory,

B--factory, SuperLEAR, high-flux electron machine, etc.\par
This abundance of projects, all studied in great detail, shows how
our field is alive. What is less encouraging,
is that most proposals  are here for years,
and we do not see any sign of serious
approval. For sure, only an ambitious project can stimulate
the community toward a long-term programme of hadronic physics in
the confinement regime. However, preliminary investigations can
be done by improving  existing machines or by using other
facilities not primarily devoted to this field.
 \subsection{Improving existing facilities}
As stressed by Vigdor \cite{Vigdor}, there are many developments
possible at Brookhaven or Fermilab, besides the main stream,
which are top quark physics and quark-gluon plasma, respectively. The same
is true for CERN. There are many workshops, study groups,
proposals, to examine the physics which can be done  apart from
LEP2 and LHC, but nothing is on the track of being approved.

SuperLEAR \cite{Amsler} is a typical example: the cost is
small, the physics programme is very
interesting, the community is active and
eager to continue, but the project is not
supported by the CERN authorities.\par
We have also been informed of projects concerning Saturne
\cite{Chamouard}. This is not the first
proposal for upgrading Saturne; I hope this
one will be further discussed.
After all , we know from the PS and AGS, that
many developments are possible, once one has
in hand a good machine.\subsection{Parasitic experiments} The
approved DA$\Phi$NE facility at Frascati will be mainly devoted to
studies of kaon decay, and in particular CP violation.  One
can also use the clean kaon beams to do strong interaction physics
\cite{Piccolo}, for instance the study of hyperons resonances, in
free air or in the nuclear medium.\par A.~d'Angelo
\cite{Annalisa} gave a status report on the GRAL project, with is
a polarized photon beam built out of the ESRF (synchrotron
radiation) at Grenoble. Again, strangeness production can be
studied, but there are many other interesting applications.

\subsection{New facilities}
We have learned that the italian community \cite{Taiuti}
is very
eager to benefit from the new facility CEBAF, which will be operative
rather soon.  Meanwhile, other electron beams
(with lower energy) will be running in
Europe. The very ambitious european project of $15\,$GeV electron
machine was not presented at this Workshop, but very present in
our mind. The name is now EEF, to recall the late EHF, and the
Workshops take place at Mainz, in the very same rooms
where the EHF proposal was elaborated.
The cost of EEF is also comparable to
that of a kaon factory. Of course, one
cannot compare the physics programme of
EEF with the huge shopping list of EHF
or KAON, which have a variety of
secondary beams, but EEF would allow
for very precise and clean investigations
of the behaviour of the quarks in the
nuclear medium. Tons of documents have
been produced, and the best QCD experts
have stressed the relevance of the
planned experiments. I think there is no
reason for postponing the decision. One
should balance the physics and the cost,
and either give the green light or
propose new means for investigating
confinement.\section*{Acknowledgments} I would like to thank once
more the organizers for the enjoyable and stimulating atmosphere of
this Workshop, and J. Cole and S. Fleck for their help in preparing the
manuscript.

\end{document}